\documentclass[11pt]{article}
\pdfoutput=1 
\usepackage{jcappub}
\usepackage{graphicx}
\usepackage{amsmath,amssymb}
\graphicspath{{./figures/}}
\usepackage{appendix}
\usepackage{placeins}
\usepackage[normalem]{ulem}
\usepackage[utf8]{inputenc}
\usepackage{pdfpages}

\input epsf
\usepackage{mathtools}
\usepackage{amsfonts}
\usepackage{upgreek}
\usepackage{latexsym}
\usepackage{stfloats}
\usepackage{afterpage}
\usepackage{tensor}
\usepackage{slashed}
\definecolor{dgreen}{rgb}{0,0.6,0}
\newcommand{\be}{\begin{equation}}
\newcommand{\ee}{\end{equation}}
\newcommand{\beq}{\begin{equation}}
\newcommand{\eeq}{\end{equation}}
\newcommand{\bea}{\begin{eqnarray}}
\newcommand{\eea}{\end{eqnarray}}

\newcommand{\bn}{{\boldsymbol{n}}}

\newcommand{\bnabla}{{\boldsymbol{\nabla}}}
\newcommand{\bee}{\boldsymbol{e}}

\newcommand{\YLM}[3]{_{#1}Y_{#2,#3}(\theta,\varphi)}
\newcommand{\YLMstar}[3]{_{#1}Y^\star_{#2,#3}(\theta,\varphi)}
\newcommand{\fsky}[0]{f_{\rm sky}}
\newcommand{\Ne}[0]{N_{\mathrm{e}}}
\newcommand{\Np}[0]{N_{\mathrm{p}}}
\newcommand{\Nc}[0]{N_{\mathrm{c}}}
\newcommand{\Ng}[0]{N_{\mathrm{g}}}
\newcommand{\zp}[0]{\zeta_{\mathrm{p}} }
\newcommand{\zc}[0]{\zeta_{\mathrm{c}} }
\newcommand{\rll}[0]{\nu_{\ell}}
\newcommand{\ii}[0]{\mathrm{i}}
\newcommand{\eem}[0]{\mathrm{e}}
\newcommand{\sd}[0]{\slashed{\partial}}
\newcommand{\sds}[0]{{\slashed{\partial}^\star}}

\newcommand{\ga}{\gamma}
\newcommand{\Ga}{\Gamma}

\newcommand{\al}{\alpha}

\newcommand{\de}{\delta}

\newcommand{\La}{\Lambda}

\newcommand{\Om}{\Omega}

\title{A new observable for cosmic shear}
\author{{Jérémie Francfort},}
\author{{Ruth Durrer}}
\author{and {Giulia Cusin}}
\affiliation{Universit\'e de Gen\`eve, D\'epartement de Physique Th\'eorique and Centre for Astroparticle Physics,
24 quai Ernest-Ansermet, CH-1211 Gen\`eve 4, Switzerland}
\affiliation{Sorbonne Université, CNRS, UMR 7095, Institut d'Astrophysique de Paris, 75014 Paris, France}

\emailAdd{jeremie.francfort@unige.ch}
\emailAdd{ruth.durrer@unige.ch}
\emailAdd{giulia.cusin@unige.ch}

\abstract{In this paper we introduce a new observable to measure cosmic shear. We show that if we can measure with good accuracy both, the orientation of a galaxy and the polarisation direction of its radio emission, the angle between them is sensitive to the foreground cosmic shear. Even if the signal-to-noise ratio for a single measurement is expected to be rather small, the fact that all galaxies in a given pixel are subject to the same shear can be used to overcome the noise. An additional advantage of this observable is that the signal is not plagued by intrinsic alignment. We estimate the SNR for the shear correlation functions $\zeta_{\pm}(\mu,z_1,z_2)$ measured in this way with the future SKA II survey.}

\begin{document}

\maketitle

\section{Introduction}
Cosmic shear is the coherent deformation of images of background galaxies due to gravitational field. It gives us precious information about the total foreground matter density as it is sensitive to both, dark and luminous matter alike.

However, shear measurements are very difficult. They typically modify the ellipticity of a galaxy by about 1\% or even less~\cite{Bartelmann:1999yn}. Furthermore, the shear correlation function is affected by so called intrinsic alignment which can be of the same order as the shear itself~\cite{Hirata:2004gc,Kirk:2010zk}. Nevertheless, in recent years several observational campains like KiDs (Kilo Degree Survey) and DES (Dark Energy Survey) and HSC (Hyper Supreme-Cam) have measured the shear  correlation function in different redshift bins, see e.g.~\cite{Aihara:2017tri,Aihara:2017paw,Hildebrandt:2018yau,KiDS:2020suj,DES:2020daw,DES:2021wwk,DES:2022qpf}. The shear correlation function is a very important variable to measure cosmological parameters and, more importantly, to  test the consistency of the cosmological standard model $\La$CDM.

The shear from scalar perturbations is determined by the lensing potential,
\be \label{eq:lensingpotential}
\phi(\bn,z) = -\int_0^{r(z)} \hspace{-2mm}\mathrm{d}r \frac{r(z)-r}{r(z)r}\left[\Phi(r\bn,t_0-r)+\Psi(r\bn,t_0-r)\right]  \,.
\ee
Here $\Phi$ and $\Psi$ are the Bardeen potentials, $r(z)$ is the comoving distance out to redshift $z$,  $t=t_0-r$ is conformal time along the light path and $\bn$ is a direction in the sky. We neglect possible contributions from tensor perturbations, i.e.\ gravitational waves, as well as from vector perturbations since they are generally small~\cite{Yamauchi:2012bc,Adamek:2015mna}.  Also, Eq.~\eqref{eq:lensingpotential} is the so called 'Born approximation', i.e. we compute the lensing potential along the straight, unlensed path, assuming that lensing is a small perturbation. For non-relativistic matter and a cosmological constant  the two Bardeen potentials are equal and correspond to the Newtonian gravitational potential.
Light from a source at redshift $z$, seen in {direction $\bn$} is coming to us from the  angular position $\bn+\bnabla\phi(\bn,z)$, where $\bnabla$  denotes the 2D gradient on the unit sphere and $\bnabla\phi(\bn,z)$ is the deflection angle.

The shear $\ga_{ij}$ is  given by the traceless part of the second  angular derivatives of $\phi$. The convergence, given by the angular Laplacian can be measured by galaxy number counts, see e.g.~\cite{Montanari:2015rga,Lepori:2021lck,Nistane:2022xuz} for theoretical aspects and numerical simulations and~\cite{Scranton:2005ci,Liu:2021gbm} for observations.

Usually, the shear is measured via the correlation of the direction of the ellipticity of galaxies. This assumes that ellipticities are intrinsically uncorrelated which is evidently not true for galaxies at similar redshifts and is even relevant for different redshifts, see e.g.~\cite{Johnston:2018nfi} for a discussion of intrinsic alignment. In this paper we derive a new observable which can be used to measure the shear correlation function and which does not depend on 'intrinsic alignment'~: 
It is well known that the polarisation of a photon is parallel transported along its path. However, a small image of finite size is Lie transported. This is in general described with the Jacobi map~\cite{Perlick:2004tq}. Therefore, if the light from a galaxy is polarised, which is usually the case for radio galaxies, and if this polarisation is aligned with the ellipticity of the galaxy, which is also typically the case, this alignment is affected by  foreground shear. Typically, the angle between the polarisation vector and the axes of the galaxy is of the order of a few degrees, see \cite{Stil:2008} for more details.
It might also be useful to measure the galaxy shapes with near future optical telescopes like LSST~\cite{LSST} or the Euclid satellite~\cite{Amendola:2016saw} but the polarisation has to be measured in the radio since these are the wavelengths of synchrotron radiation whose polarisation is correlated with the intrinsic direction of the galaxy.

If the principle axes of the shear tensor and the intrinsic ellipticity of the galaxy are not aligned, this leads to a slight rotation of the image with respect to the polarisation, as we have shown in a previous paper~\cite{Francfort:2021oog}. In that paper we have studied the effect considering galaxies as Schwarzschild lenses. In this work, we use shear from linear cosmological perturbation theory and want to outline how one can use the correlation of the orientation of the image and the polarisation to measure the shear correlation function.
The class of sources we have in mind in this analysis are low frequency radio
galaxies (typically 1-50 GHz as lower frequencies are significantly depolarised by Faraday
rotation \cite{Mahatma_2020}), for which the dominant source of linear polarisation is expected to be
synchrotron radiation due to electrons moving in the magnetic field of the galaxy. For
these objects, the magnetic field is dominantly in the galactic plane (the
orthogonal component is very small) and tends to be aligned with galaxy morphology,
i.e. the semi-major axis of the galaxy (see e.g. \cite{ 1966ARA&A...4..245G}). Then polarisation from synchrotron
radiation is mainly orthogonal
to the magnetic field component (i.e. it is in the orbital plane). Hence its projected component (on the observer's screen) is normal to the galaxy’s
major axis.

Previous authors have exploited the fact that the polarisation position angle is unaffected by lensing in order to measure gravitational lensing of distant quasars, see~\cite{1991ApJ...367L...1K, kronberg1996estimates, Burns_2004}. In \cite{Brown_2010}, the authors proposed to use the polarisation information in radio galaxies as an indicator of the galaxy intrinsic shape, with the goal of mitigating shot noise and intrinsic alignement uncertainties in shear reconstruction. In \cite{Brown_2011}, the same authors extended this idea to reconstruct maps of the projected dark matter distribution, or the lensing convergence field. The authors of  \cite{Whittaker:2015fma} proposed to use a proxy for the intrinsic position angle of an observed galaxy, and propose  techniques for cleanly separating weak gravitational lensing signals from intrinsic alignment contamination in forthcoming radio surveys. Finally, in \cite{Camera:2016owj} it is shown that, thanks to polarisation information, radio weak lensing surveys will be able to mitigate contamination by intrinsic alignments, in a way similar but fully complementary to available self-calibration methods based on position-shear correlations. 

Unlike all these works, where the polarisation direction is used to have a better handle on intrinsic alignment (inferred from the polarisation direction itself), we propose to measure the offset between the observed polarisation and galaxy morphology as a new observable on its own. 
In other words, although the idea of using polarisation information to access the galaxy intrinsic orientation is widely explored around in the literature, we believe that this is the first time where a shear estimator is explicitly written down in terms of the offset between the (observed) galaxy major axis and polarisation orientation. A first attempt to do weak lensing with radio surveys is published in~\cite{Harrison:2020zsv}. In this first work, however polarisation is not used.

In Ref.~\cite{Whittaker:2017hnz} the authors do consider rotation but not the rotation induced by shear which is considered in the present paper, rather they consider the rotation from an antisymmetric contribution to the Jacobi map which is much smaller than shear as it appears only at second order in the perturbations~\cite{Fanizza:2022wob}.

This paper is structured as follows. In the next section we develop the theoretical expressions which determine the shear from a measured angle $\de\al$ by which the orientation of the galaxy and its polarisation differ. In Section~\ref{s:error} we present a rough estimate of the error on the measurement given a typical precision of measured angles. In Section~\ref{s:res} we discuss our results and in Section~\ref{s:con} we conclude. Some useful properties of Spin Weighted Spherical Harmonics are presented in Appendix~\ref{app:spefun} for completeness. In Appendix~\ref{app:error} we derive in detail the error estimates used in the main text.

\vspace{0.5 cm}
\noindent \textbf{Notations and conventions:}\\
We use the signature $(-,+,+,+)$.\\
The usual spherical angles are $(\theta, \varphi)$, and the corresponding unit vector is $\boldsymbol n$. The surface element of the sphere is denoted $\mathrm d \Omega$.
The lensing potential is called $\phi(\boldsymbol n,z)$.
The Bardeen potentials are $\Phi$ and $\Psi$.
The Spin Weighted Spherical Harmonics are  $_sY_{\ell,m}$, while the 'usual' Spherical Harmonics, $_0Y_{\ell,m}$, are simply denoted $Y_{\ell,m}$.

\begin{figure}[ht!]
\begin{center}
\includegraphics[width=0.5\textwidth]{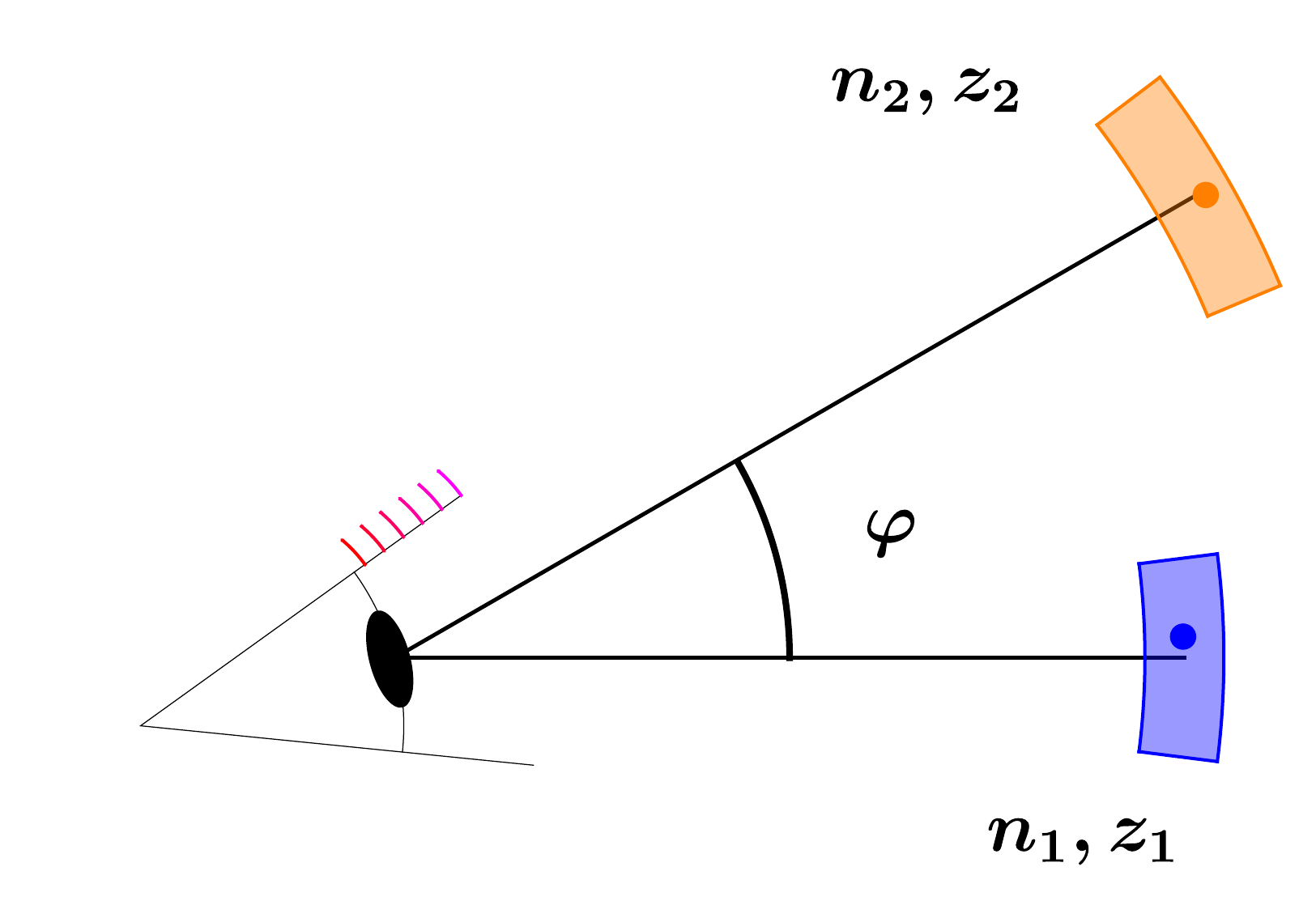}  
\end{center}
\caption{\label{fig:setup} 
The general setup (in $2D$, seen from above): Two pixels are considered, in directions $\boldsymbol{n}_j$ and at redshifts $z_j$. The directions are separated by an angle $\varphi$ (or $\boldsymbol{n}_1 \cdot \boldsymbol{n}_2 = \cos \varphi = \mu$). In each pixel, one galaxy is chosen (represented by the dots). The computations are made in the equatorial plane.}
\end{figure} 
    
\section{Theoretical development}\label{s:theo}
The lensing potential given in Eq.~\eqref{eq:lensingpotential} is a stochastic quantity which can be decomposed into Spherical Harmonics as
\be \label{eq:lensingdec}
\phi(\boldsymbol n,z) = \sum_{\ell,m} \phi_{\ell,m}(z) Y_{\ell,m}(\boldsymbol n)\,,
\ee
where the scalars $\phi_{\ell,m}(z)$ are also random variables.
Assuming statistical isotropy different values of $\ell$ and $m$ are not correlated and  their two-point correlation spectrum is given by
\be
\langle \phi_{\ell_1, m_1}(z_1) \phi^*_{\ell_2,m_2}(z_2) \rangle
= C_{\ell_1}(z_1,z_2) \delta_{\ell_1, \ell_2} \delta_{m_1,m_2}\,.
\ee 
The $C_{\ell}(z_1,z_2)$ are the lensing power spectra for different resdhifts $z_1$ and $z_2$.  If the fluctuations are Gaussian, these power spectra encode all the statistical information of the lensing potential. The lensing potential contains very useful information e.g. about the matter distribution in the Universe which is not plagued by the biasing problem of galaxy number counts. Therefore estimating it using different measurements with different systematics is very important. 

In this section we present the main theoretical tools and formulas of the article. More explanations and details can be found in the Appendix.

We consider radio galaxies which are polarised along their semi-major (or minor) axis. This polarisation is parallel transported and hence its components expressed in a parallel transported Sachs basis are constant. The radio galaxy, represented by an ellipse is sheared and magnified according to the Jacobi map. If the principle axes of the shear are not aligned with the principle axes of the  galaxy, this leads to a rotation of the galaxies principle axes expressed in the Sachs basis.  In our previous work \cite{Francfort:2021oog} we have calculated this rotation which is given by
\be\label{e:deal}
\de\al =  \frac{\varepsilon^2}{2-\varepsilon^2}\left(\gamma_2 \cos 2 \alpha - \gamma_1 \sin 2 \alpha\right)\,.
\ee
Here $\varepsilon$ is the eccentricity of the galaxy, $(\ga_1,\ga_2)$ are the components of the shear matrix in the Sachs basis,
\be
\boldsymbol{\Ga} = \left(\begin{array}{cc}-\ga_1 & -\ga_2\\ -\ga_2 & +\ga_1\end{array}\right) \,,
\ee
and $\alpha$ is the angle between the major-axis of the galaxy shape and the first basis vector $\bee_1$.
We stress that the dependence of the rotation angle \eqref{e:deal} on the choice of the Sachs basis is only apparent: under a rotation of the Sachs basis, the shear transformation compensates the transformation of the position angle $\alpha$, see \cite{Francfort:2021oog} for details.

If the semi-major axis of the galaxy is aligned with the shear, $\de\al$ vanishes. For example if we choose $\bee_1$ in the direction of the semi-major axis of the galaxy such that $\al=0$, alignment with the shear implies $\ga_2=0$ and hence $\de\al=0$. In this case, the shear just enhances or reduces somewhat the ellipticity of the galaxy. In all other situation it generates also a rotation by $\de\al$. This rotation has already been studied long ago as a possible origon of the anisotropy of galaxy orientations~\cite{Par}.

An additional rotation is in principle also generated by the anti-symmetric part of the Jacobi matrix. But this part in non-vanishing only at second order in perturbation theory~\cite{Fanizza:2022wob} and we neglect it here.

In addition to $\de\al$, the angle between the polarisation direction and the semi major  axis, also the eccentricity $\varepsilon$ and the direction of the galaxy's semi-major  axis parametrised by $\al$ are observables. Similar to our previous work~\cite{Francfort:2021oog}, we define an observable which we call the 'scaled rotation'\footnote{Some versions of the article have a sign mistake in this definition. The formula given here is correct.} by
\be 
\Theta = \frac{2-\varepsilon^2}{\varepsilon^2} \delta \alpha\,.
\ee
With \eqref{e:deal} the scaled rotation is related to the shear as
\be \label{eq:scaledrot}
\Theta = \gamma_2 \cos 2 \alpha - \gamma_1 \sin 2 \alpha\,,
\ee
which is actually simply the shear in the direction $\al-\pi/4$, see Appendix~\ref{sec:invcorfun}. We want to determine the correlation function $\langle\Theta(\bn_1,z_1)\Theta (\bn_2,z_2)\rangle $ for two directions $\bn_1$ and $\bn_2$ in the sky and two redshifts $z_1$, $z_2$. Our expression for the variable $\Theta(\bn,z)$ given in \eqref{eq:scaledrot} in principle depends on our choice for the Sachs basis via the angle $\alpha$ and via $\ga_1$ and $\ga_2$. However, as explained in App.~\ref{sec:invcorfun}, one can circumvent this problem and define a correlation function that is explicitly coordinate invariant by choosing $\boldsymbol{e}_1$ the direction of the great circle from $\bn_1$ to $\bn_2$ which is equivalent to putting both galaxies on the 'Equator', with  coordinates $(\pi/2,0)$ and $(\pi/2, \varphi)$ and  $\mu=\cos\varphi=\bn_1\cdot\bn_2$. Note that there is still a $\mathbb{Z}_2$ symmetry where one can swap both galaxies. However, the correlation function does not depend on this choice. Given two galaxies and the described setup, the correlation between their scaled rotation is given by
\begin{align} \label{eq:corrtheta}
    \langle {\Theta}(\boldsymbol{n}_1,\alpha_1,z_1) {\Theta}(\boldsymbol{n}_2,\alpha_2,z_2) \rangle
    &= \zeta_+(\mu,z_1,z_2) \cos(2(\alpha_1- \alpha_2)) + \zeta_-(\mu,z_1,z_2) \cos(2(\alpha_1+\alpha_2))\,,
\end{align}
with $\mu = \boldsymbol n_1 \cdot \boldsymbol n_2 = \cos \varphi$ and  $\zeta_+(\mu,z_1,z_2)$ and $\zeta_-(\mu,z_1,z_2)$ the two coordinate independent shear correlation functions (see App.\,\ref{app:spefun} for more details). These correlation functions are related to the power spectrum of the lensing potential $C_{\ell}(z_1,z_2)$ as 
\begin{align}
    \label{eq:zetapexpri}
    \int_{-1}^{+1} \zeta_+(\mu,z_1,z_2) \tilde{P}_\ell(\mu)\, \mathrm{d}\mu &= \frac{1}{4\pi} C_\ell(z_1,z_2) \rll^2 \,, \\
    \label{eq:zetamexpri}
    \int_{-1}^{+1} \zeta_-(\mu,z_1,z_2) \tilde{Q}_\ell(\mu)\, \mathrm{d}\mu &= \frac{1}{4\pi} C_\ell(z_1,z_2) \rll^2 \,,
    \\
    \nu^2_\ell & = \frac{(\ell+2)!}{(\ell-2)!}\,,
\end{align}
where the polynomials $\tilde{P}_\ell(\mu)$ $\tilde{Q}_\ell(\mu)$ are defined by, $\mu=\cos\theta$,
\begin{align}
        - \sqrt{\frac{2\ell+1}{4\pi}}\; _{+2}\tensor{Y}{_{\ell,+2}}(\theta,\pi/2) &=
        - \sqrt{\frac{2\ell+1}{4\pi}}\;
        {_{-2}\tensor{Y}{_{\ell,-2}}(\theta,\pi/2)} 
        =\frac{2\ell+1}{16\pi} \tilde{Q}_{\ell}(\mu) \,, \\
        - \sqrt{\frac{2\ell+1}{4\pi}}\; _{+2}\tensor{Y}{_{\ell,-2}}(\theta,\pi/2) &=
        - \sqrt{\frac{2\ell+1}{4\pi}}\;
        {_{+2}\tensor{Y}{_{\ell,-2}}}(\theta, \pi/2)
        =\frac{2\ell+1}{16\pi} \tilde{P}_{\ell}(\mu) \,, 
\end{align}
and $ _sY_{\ell,m}$ are the Spin  Weighted Spherical Harmonics. More details, and the explicit expressions for $\ell=2,\dots,5$ are given in App.~\ref{app:spefun}.

From the observable $\Theta$ we now  construct  an estimator for the coordinate independent correlation functions $\zeta_+$ and $\zeta_-$. Since we want to estimate two correlation functions, we need two couples of galaxies, separated by the same angle $\varphi$. Schematically, as $\Theta \sim \gamma_1 + \gamma_2$, one needs two galaxies to invert this relation and express $\gamma_1$ and $\gamma_2$. Moreover, as the correlation function is given by $\zeta \sim \langle \gamma \gamma\rangle $, we need the value of $\gamma$ in two different pixels, which can be performed considering $4$ galaxies in total.

More precisely, the estimator can be computed as follows. We consider two couples of galaxies both separated by the same angle $\varphi$ and located at the same redshifts (within the resolution of our survey). The galaxies of the first couple  have the directions and redshifts  $(\boldsymbol n_j,z_j)$ with angles as defined above $\alpha_j$ ($j=1,2$), while the second couple of galaxies are located in different directions $\boldsymbol n^\prime_j$ and with different angles $\alpha^\prime_j$ but inside the same redshift bins $z_j$. Note that we define the angles $\alpha_j$ and $\alpha^\prime_j$, with respect to the great circle connecting $\bn_1$ and $\bn_2$ respectively $\bn'_1$ and $\bn'_2$ which can be different for each couples. The two couples of galaxies, however should be separated by the same angle $\varphi$ (within our angular resolution), i.e. $\boldsymbol n_1 \cdot \boldsymbol n_2 = \boldsymbol n^\prime_1 \cdot \boldsymbol n^\prime_2 = \cos \varphi = \mu$. The two observables are the product of the scaled rotations, namely
\begin{align}
\Xi &= {\Theta}(\boldsymbol n_1,\alpha_1,z_1)  \Theta(\boldsymbol n_2, \alpha_2,z_2)\,,  \\
\Xi^\prime &= {\Theta}(\boldsymbol n_1^\prime, \alpha_1^\prime,z_1)  \Theta(\boldsymbol n_2^\prime, \alpha_2^\prime,z_2)\,. 
\end{align}
From these, and using the theoretical expression of the correlation function of the scaled rotations given by Eq.~\eqref{eq:corrtheta}, replacing the expectation value by the observables $\Xi$ and $\Xi^\prime$, we can extract the estimators
\begin{align} 
        \label{eq:zetapest}
        \hat{\zeta}_+(\mu,z_1,z_2) &= \Xi\, F_1(\alpha_1^\prime, \alpha_2^\prime, \alpha_1, \alpha_2) 
        + \Xi^\prime\, F_1(\alpha_1, \alpha_2,\alpha_1^\prime, \alpha_2^\prime) \,, \\
        \label{eq:zetamest}
        \hat{\zeta}_-(\mu,z_1,z_2) &= \Xi\, F_2(\alpha_1^\prime, \alpha_2^\prime, \alpha_1, \alpha_2) 
        + \Xi^\prime\, F_2(\alpha_1, \alpha_2,\alpha_1^\prime, \alpha_2^\prime) \,, 
\end{align}
with 
\be\label{e:F1}
F_1(\alpha_1, \alpha_2, \alpha_1^\prime, \alpha_2^\prime) =
\frac{\cos (2 (\alpha_1 + \alpha_2))}{ \cos(2(\alpha_1^\prime-\alpha_2^\prime)) \cos(2(\alpha_1+\alpha_2)) - \cos(2(\alpha_1^\prime+ \alpha_2^\prime)) \cos(2(\alpha_1-\alpha_2)) } \,,
\ee
and
\be \label{e:F2}
F_2(\alpha_1, \alpha_2, \alpha_1^\prime, \alpha_2^\prime) =
\frac{\cos (2 (\alpha_1 - \alpha_2))}{ \cos(2(\alpha_1^\prime+\alpha_2^\prime)) \cos(2(\alpha_1-\alpha_2)) - \cos(2(\alpha_1^\prime- \alpha_2^\prime)) \cos(2(\alpha_1+\alpha_2)) } \,.
\ee
{We observe that in eqs.\,(\ref{eq:zetapest}) and (\ref{eq:zetamest}) on the left hand side there is no angle dependence. We used this notation to stress that, observationally, one chooses a given Sachs frame and for each galaxy quadruplet in pixels $\bn_1$ and $\bn_2$, one builds the correlations given in eqs.~(\ref{eq:zetapest}) and (\ref{eq:zetamest}). Every single estimator depends on the frame choice. However, their expectation value obtained by averaging over all possible quadruplets in the two pixels is independent of the angles $\al_i$ and $\al'_i$. In other words, and by construction,
\be
 \langle\hat{\zeta}_{\pm}(\mu,z_1,z_2)\rangle = \zeta_{\pm}(\mu,z_1,z_2)\,.
\ee

Once an estimator for $\zeta_{\pm}$ is obtained, the estimator for the lensing potential power spectrum $C_\ell(z_1,z_2)$ can be given by  Eqs.~\eqref{eq:zetapexpri} and \eqref{eq:zetamexpri}.

\section{Error estimation}\label{s:error}
In this Section, we estimate the expected error (or signal-to-noise ratio) on the lensing angular power spectrum extracted via Eqs.~\eqref{eq:zetapexpri} and \eqref{eq:zetamexpri}, starting from our estimator for the correlation functions  Eqs.~\eqref{eq:zetapest} and \eqref{eq:zetamest}. 

As explained in the previous section, given two couples of galaxies, each couple being separated by an angle $\varphi$ (with $\mu = \cos \varphi$), an estimator for the correlation functions $\zeta_{\pm}$ is given by Eq.~\eqref{eq:zetapest} and Eq.~\eqref{eq:zetamest}.
Of course, to obtain a good estimator for $\zeta_{\pm}(\mu,z_1,z_2)$ we need to have many pairs of galaxies at a given angular separations $\varphi$ (with $\mu=\cos\varphi$) inside the two redshift bins. 
Furthermore, we need a good measurement of the scaled rotation for these pairs and a good measurement of the angles $\al_j$ and $\al_j'$. The expressions for $F_1$ and $F_2$ (see Eqs.\,(\ref{e:F1}) and (\ref{e:F2})) also tell us that for $\al_1+\al_2 = \al'_1+\al'_2 = \pi/4$ we cannot determine $\hat\zeta_+$ while for for $\al_1-\al_2 = \al'_1-\al'_2 = \pi/4$ we cannot determine $\hat\zeta_-$. It follows that to obtain a well-defined estimator of the correlation functions $\zeta_{\pm}$ we need to select properly the angles $\al_j$ and $\al_j'$, excluding galaxy pairs with $\al_1+\al_2 = \al'_1+\al'_2 = \pi/4$ or with  $\al_1-\al_2 = \al'_1-\al'_2 = \pi/4$. 
Note, however, it does not matter whether the angles $\al_j,~\al_j'$ are correlated, hence intrinsic alignment, the major concern for traditional shear measurements is not an issue here. What is important, however, is to have a good measurement of these angles and of the small and more difficult-to-measure angle $\de\al$ between the image axis and polarisation.

An optimal estimator can be built as explained in Appendix \ref{app:error}, by combining the information that can be extracted from all possible pairs of couples with the same angular separation and redshifts. It is optimal to choose the weighting of each measurement inversely proportional to its error. To determine the associated signal-to-noise ratio (SNR), we use the results presented in Appendix~\ref{app:error}. Let $q$ represent a pair of a couples of galaxies (hence a quadruplet). For each $q$, we compute an estimator $\hat{\zeta}_{\pm, q}(\mu)$ with its relative error $\tau_{\pm,q}$. The total signal-to-noise ratio for the measurement of $\zeta_{\pm}(\mu,z_1,z_2)$ is given by Eq.~\eqref{eq:SNR}
\be
{\rm SNR}_{\pm}(\mu,z_1,z_2)= 
\sqrt{\sum_q \frac{1}{\tau^2_{\pm,q}}} \,.
\ee
This sum can be computed explicitly if one is given a catalogue of measurements. Here, we will take a more heuristic approach and admit that the relative error is roughly equal (or we just consider an average value)
\be
\tau_{\pm, q} \simeq \tau_0\,.
\ee 
Then, the signal-to-noise is estimated as
\be \label{eq:SNRfinal}
{\rm SNR}_{\pm}(\mu,z_1,z_2) \approx \frac{\sqrt{\Ne(\mu,z_1,z_2)}}{\tau_0}\,,
\ee 
where $\Ne(\mu,z_1,z_2)$ is the number of estimators one can extract by choosing two couples of galaxies separated by an angle $\varphi$. The number of quadruplets is computed in  Appendix~\ref{app:error} and is given by 
\be \label{eq:nevalue}
\Ne(\varphi,z_1,z_2) = 
\frac{\Np(\varphi,z_1,z_2) (\Np(\varphi,z_1,z_2)-1)}{2} 
\simeq \frac{1}{2}\left(\Ng(z_1) \Ng(z_2) \frac{8 \fsky \sin \varphi}{\delta \theta^3}\right)^2\,,
\ee 
where $\Ng(z)$ is the number of galaxies in a pixel at redshift $z$ and $\delta \theta$ is the aperture of the angular resolution. Note that the formula for $\Ne$ given here holds for two different redshifts, and has to be divided by $4$ if the considered redshifts are equal.

The final result for the signal-to-noise ratio given by Eq.~\eqref{eq:SNRfinal} shows that even if the erorr on a single estimator is typically rather large so that  $\tau_0 >1$, the quality of the best estimator can still be good if we have sufficiently many galaxies at our disposal.

Note that here, we assumed that all the individual estimators are statistically independent. In reality, this is not the case, as we can assume that the galaxies in the same pixel are somehow correlated (either their shape or their orientation). Hence intrinsic alignment enters here in the error estimate but not in the signal. Furthermore, in the number of estimators given in \eqref{eq:nevalue} the same couples of pixels are used multiple times. We therefore prefer to use a more pessimistic estimation for the number of independent estimators  setting 
\be  \label{eq:Necon}
\Ne(\varphi,z_1,z_2) \simeq N_c(\varphi) = \frac{8 \fsky \sin \varphi}{\delta \theta^3}\,,
\ee
where $\Nc$ is the number of couples of pixels separated by an angle $\varphi$. Here we admit just one galaxy from each pixel. More details can be found in Appendix~\ref{app:error}.

Finally, and to conclude this Section, another method would be to simply compute the estimated shear field ${\boldsymbol{\gamma}}(\boldsymbol{n},z)$ in every pixel using Eq.~\eqref{eq:scaledrot}. By doing this, the signal-to-noise ratio for every pixel would be given by $\sqrt{\Ng}/\tau_0$, where $\Ng$ is the galaxy number in this specific pixel and $\tau_0$ is the mean relative error on one measurement. In this way one could construct a shear map  in the sky for each redshift bin. From this map one can then extract the power spectrum with its associated error. As we know e.g. from CMB lensing maps~\cite{Planck:2018lbu}, even if the map itself is noise dominated, we can obtain a good estimator for its power spectrum. \textcolor{red}{}Note that to extract the shear in one pixel, one needs to consider only a pair of galaxy, as the shear has two real components $\gamma_1$ and $\gamma_1$. However, to compute the shear correlation function, one needs to know the shear in two pixels. In other words, even in this context, it is necessary to have two pairs of galaxies to build an estimator for the correlation function. The selling argument for the method we present here is that one could, in principle, construct a map of the cosmic shear simply considering pairs of galaxies, without taking into account a potential intrinsic correlation.}

\section{Results and discussion}\label{s:res}
\begin{figure}[ht!]
\begin{center}
\includegraphics[width=0.47\textwidth]{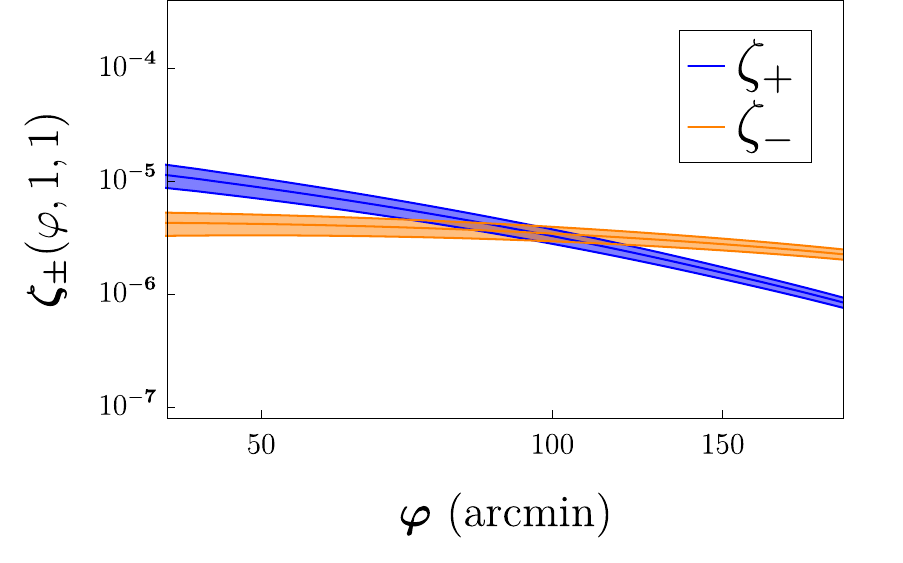}  
\includegraphics[width=0.47\textwidth]{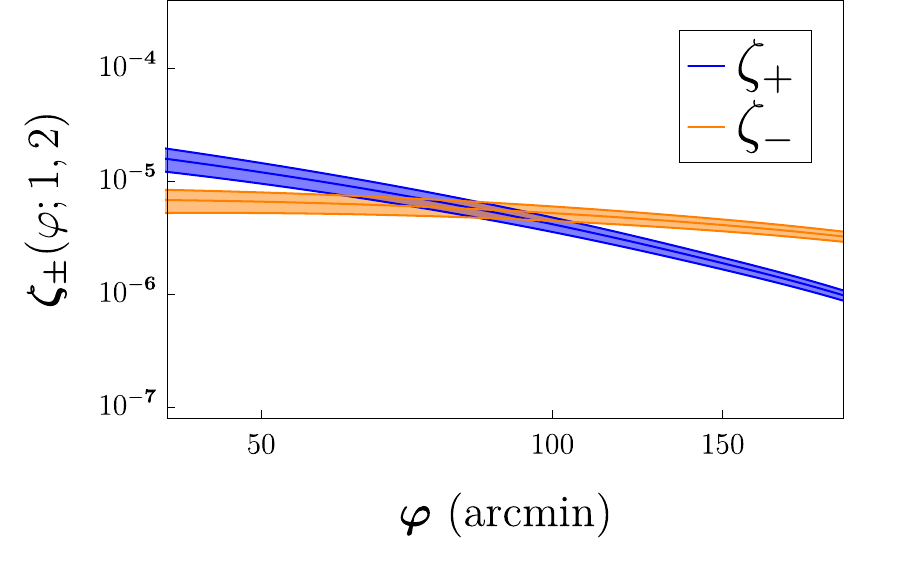}  
\includegraphics[width=0.5\textwidth]{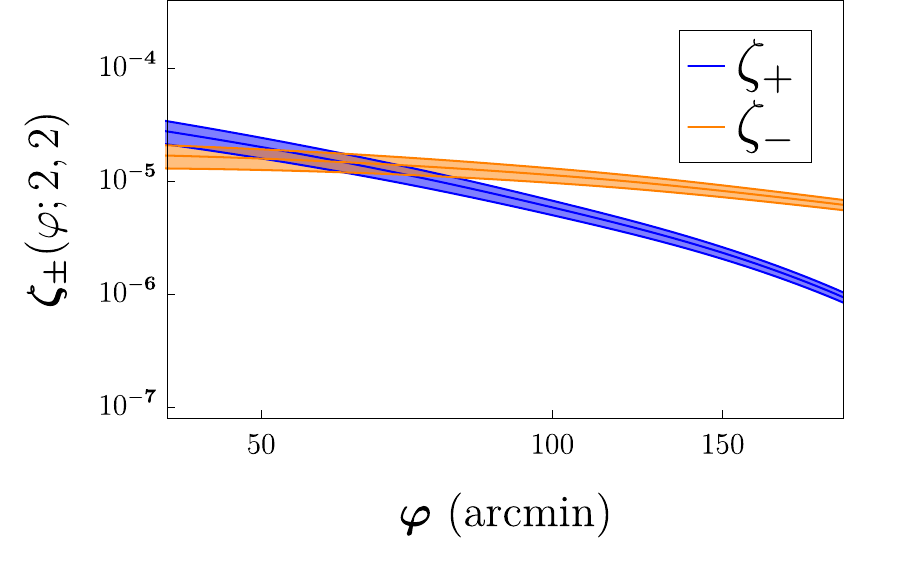}  
\end{center}
\caption{\label{fig:zetapm11} 
Correlation functions $\zeta_\pm(\varphi)$ computed with the flat-sky approximation. The  $C_\ell$ are taken from CLASS, and the sum is truncated at $\ell=20'000$. The error bars were computed with $\mathrm{SNR}= 40 \sqrt{\sin \varphi}$. We consider redshift bins $(z_1, z_2)=(1,1), (1,2)$ and $(2,2)$ (from top left to bottom panel).}
\end{figure}

In Fig.~\ref{fig:zetapm11}, we show an example of the results we can obtain. As discussed in the previous section, we assume $\Ng=1$ to take into account that the galaxies in the same pixel are not independent from each other, and use Eq.~\eqref{eq:Necon}. The parameters are taken from SKA2, see \cite{Bull:2016}\ for more details. We choose a sky fraction and a pixel size of
\begin{align}
    \fsky &\approx 0.7\,, \\
    \delta \theta &= 5^\prime \approx 1.4\times 10^{-3}\,.
\end{align}
Moreover, the typical shear signal $\ga$ will be of order $10^{-3}$. For a precise estimate of the error per galaxy pair, we would need precise values for the errors on the various quantities as they are available once a mission is planned. To get a pessimistic rough estimate, we have realised several simulation using an error of $\pi/5$ on the angles and $1/2$ on $\varepsilon$. This leads to a conservative relative error per galaxy pair of the order of $\tau_0\approx 10^{3}$. This estimate is pessimistic, as in real experiments one can hope to make this error smaller. On the other hand, the assumption that the polarisation is perfectly aligned with the main axes of the galaxy is optimistic. The idea is that these two assumptions might roughly compensate each other,leading to the right order of magnitude for the resulting estimate. Of course this treatment is simplistic and for a real observational campaign, detailed simulations will be necessary. Inserting these numbers in \eqref{eq:Necon} and \eqref{eq:SNRfinal} we obtain a signal-to-noise ratio of order \be 
{\rm SNR} \approx 45 \sqrt{\sin \varphi}\,.  
\ee 
This is the signal-to-noise ratio for our estimator $\hat\zeta_\pm(\varphi,z_1,z_2)$ in two redshift bins around $z_1$ and $z_2$ and within one angular bin. One also needs the estimated value of $\zeta_\pm(\varphi)$, which would be 
obtained from a catalogue with the method we describe in this paper. As we do not  yet have such a catalogue, we compute the theoretical value of the correlation function. We compute the  power spectrum of the lensing potential, $C^\phi_\ell(z_1,z_2)$ for $(z_1,z_2)=(1,1), (1,2), (2,2)$ with CLASS~\cite{Lesgourgues:2011re,Blas:2011rf} using the by default parameters from the Planck 2018
data~\cite{Planck:2018vy} ($h=0.6781\,, 
    h^2\Om_{\mathrm{cdm}}= 0.0552278 \,, 
    h^2\Om_{\mathrm b}=  0.0102921 \,, 
    \log(10^{9}A_{\mathrm s})= 0.742199 \,,
    n_{\mathrm s} = 0.9660499$)
 To compute the correlation functions, one would need to invert the relations \eqref{eq:zetapexpri} \eqref{eq:zetamexpri}, i.e. evaluate the sums Eq.~\eqref{eq:zpexpr} and Eq.~\eqref{eq:zcexpr}. However, the polynomials $\tilde{P}$ and $\tilde{Q}$ are highly oscillating as $\ell$ gets large and the computation is very badly converging. Instead, we use the flat sky approximation, see \cite{Kilbinger:2014cea} and \cite{Bartelmann:2010fz} for more details, to approximate the correlation functions as
\bea 
\zeta_+(z_1,z_2,\varphi) &=& \frac{1}{2\pi} \int_0^{\infty}\; \ell J_0(\ell \varphi) \frac{1}{4} \left(\ell(\ell+1)\right)^2  C_\ell(z_1,z_2)\, \mathrm{d}\ell\,\\
\zeta_-(z_1,z_2,\varphi) &=& \frac{1}{2\pi} \int_0^{\infty}\; \ell J_4(\ell \varphi) \frac{1}{4} \left(\ell(\ell+1)\right)^2  C_\ell(z_1,z_2)\, \mathrm{d}\ell\,.
\eea 
Truncating the integral at $\ell=20'000$ seems reasonable, as the relative error is less than $10^{-3}$ in this case, which is much smaller than the inverse signal-to-noise ratio. 

In Fig.~\ref{fig:zetapm11} we show the results for the correlation functions $\zeta_\pm(\varphi)$ computed in the flat-sky approximation. The shaded region around each curve represents the uncertainty computed with ${\rm SNR}= 40 \sqrt{\sin \varphi}$. Different panels correspond to different redshift bins. The result is not very sensitive to the thickness of the redshift bins. In a true survey this is an advantage as it allows us the enhance the number of galaxies per bin.

\section{Conclusions}\label{s:con}
In this paper we proposed a new method to extract the shear correlation function, by measuring the correlation function of the angle between the  image major axis and the polarisation direction of radio galaxies. 
In particular, we built an estimator for the shear correlation function given two couples of galaxies separated by an angle $\varphi$, and estimated the error one gets by combining all possible pairs separated by this angle.

The advantage of this method with respect to traditional shear measurements is that we do not rely on the assumption that galaxy  eccentricities are uncorrelated, hence we do not have to deal with a  parametrisation of intrinsic alignment and its uncertainties, which are one of the major source of error in standard shear measurements in present and planned surveys~\cite{Hirata:2004gc,Kirk:2010zk,Hildebrandt:2018yau,KiDS:2020suj,DES:2020daw,DES:2022qpf}.Even though our signal does not depend on intrinsic alignment, we have seen that the error does since intrinsic alignment correlates the measurements from different galaxies which therefore cannot be considered as independent estimators.
In the presented estimation of the signal-to-noise we have taken this into account in a very conservative way, assuming that we can make only 1 independent measurement per pixel.

We find that even if the signal-to-noise ratio for a single measurements (i.e. for a given galaxy quadruplet) is expected to be rather small, the fact that all galaxies in a given pixel are subject to the same shear can be used to overcome the noise. As a case study, we considered the specifications of SKA2: the number of independent estimators for a given angular separation $\varphi$ and two redshifts $z_1$, $z_2$ is expected to scale as $\sim 10^9 \sin\varphi$. As a consequence, the noise on a single measurement can exceed the signal by a factor $10^3$, and still yield an signal-to-noise of order 40 which is largely sufficient to detect the signal. 
Therefore, even if the maps of $\delta\alpha$ measurements for each redshift bin will be largely noise dominated, we will be able to obtain a good estimate of for the shear correlation function when
combining all the measurements together.  

We stress that the goal of the present paper was to present a new method to reconstruct the shear correlation functions with a new observable, and to build an estimator for it.
Of course, the limiting factor of our forecasts is that we had to assume some number for the precision with which the various angles $\delta \alpha$ and $\alpha$ can be measured. However, as explained above, our choice of errors is quite conservative, and the crucial factor setting the signal-to-noise level of our estimator is the high statistics.  For this reason, we do not expect a more refined analysis to drastically change the conclusions of our study. 

Finally we point out that, while in this work we focused on the reconstruction of the shear correlation function, our new observable can be used also to get a shear sky map. This is another advantage of our method with respect to standard shear reconstruction methods, which look at galaxy shapes only (from the study of galaxy ellipticity it is not possible to get a shear mapping, but only to extract correlation functions). 
 A natural extension of our work is to apply this  method to simulated (or real) galaxy lensing and polarisation data. This would provide us with a more realistic estimate of the uncertainties, and allow us to compare this shear reconstruction method with traditional LSST/Euclid techniques to measure the shear correlation function.

\section*{Acknowledgements}
We thank Richard Battye, Michael Brown, Charles Dalang, Ian Harrison, Alan Heavens, Azadeh Moradinezhad Dizgah, Serge Parnovskii, Cyril Pitrou and Isaac Tutusaus for useful discussions and comments.\\
We are very grateful to Francesca Lepori for her valuable help with {\sc class}.\\
This work is supported by the Swiss National Science Foundation.

\appendix
\section{Special functions} \label{app:spefun}
\subsection{Spin Weighted Spherical Harmonics}
\label{sec:swsh}
This appendix follows Refs.~\cite{Durrer:2020fza,Seibert:2018}.
Let $(\theta, \varphi)$ be the usual spherical coordinates on the sphere. We define the Spin Weighted Spherical Harmonics, $ _sY_{\ell,m}$, where $s$ represents the weight. The $s=0$ Spherical Harmonics are the usual Spherical Harmonics functions $\tensor[_0]{Y}{}_{\ell,m}\equiv\tensor{Y}{}_{\ell,m}$, with the convention
\be 
Y_{\ell,-m} = (-1)^m Y_{\ell,m}^\star\,.
\ee
For a generic integer $s$, we define first the spin raising and spin lowering operations, $\sd $ and $\sd ^\star$,  on a  function $\tensor[_s]{Y}{}_{\ell,m}$ with spin weight $s$ as
\begin{align} \label{eq:sd}
\sd _sY_{\ell,m} &=
\left( s \cot \theta - \partial_\theta - \frac{\ii}{\sin \theta} \partial_\varphi \right) \,_sY_{\ell,m}  \,,
\end{align}
and
\begin{align} \label{eq:sds}
\sds  _sY_{\ell,m}  &=
{\left( -s \cot \theta - \partial_\theta + \frac{\ii}{\sin \theta} \partial_\varphi \right)} \,_sY_{\ell,m} \,.
\end{align}
The Spin Weighted Spherical Harmonics for generic $s\in \mathbb{Z}$ are obtained recursively with the spin raising and spin lowering operators given by Eq.~\eqref{eq:sd} and Eq.~\eqref{eq:sds} via
\begin{align}
\sd _sY_{\ell,m}   &= \sqrt{(\ell-s)(\ell+s+1)} \; 
 _{s+1}Y_{\ell,m} \,,  \\
\sds _sY_{\ell,m}  &=- \sqrt{(\ell+s)(\ell-s+1)}\; 
_{s-1}Y_{\ell,m} \,,
\end{align}
together with the starting point $_0\tensor{Y}{}_{\ell,m}\equiv\tensor{Y}{}_{\ell,m}$. Hence, the slashed derivatives can be interpreted as spin raising/lowering operators.
In particular, for $s=\pm 2$, these definitions yield
\begin{align} \label{eq:ytwo}
\sd^2{Y}{_{\ell,m}}   &= \rll \; \;  {_{2}\tensor{Y}{_{\ell,m}}}\,,  \\
\sds^2 {Y}{_{\ell,m}}   &= \rll\;\;  {_{-2}\tensor{Y}{_{\ell,m}}}\,, \\
\mbox{where}~~\rll &= \sqrt{\frac{(\ell+2)!}{(\ell-2)!}}\,.
\end{align}
The Spin Weighted Spherical Harmonics satisfy the orthogonality condition ( $d\Omega=\sin\theta d\theta d\varphi$)
\be \label{eq:yortho}
\int\; \YLM{s}{\ell_1}{m_1} \, \YLMstar{s}{\ell_2}{m_2}  \, \mathrm d \Omega = \delta_{\ell_1,\ell_2} \delta_{m_1,m_2}\,,
\ee
and the conjugation relation 
\be
\label{eq:ylmconj}
_{-s}Y_{\ell,-m} = (-1)^{s+m}\,\YLMstar{s}{\ell}{m}\,.
\ee 
The Spin Weighted Spherical Harmonics also satisfy the the following addition theorem
\be \label{eq:summation}
\sqrt{\frac{4\pi}{2\ell+1}} 
\sum_m  \;
{_{s_1}\tensor{Y}{_{\ell,m}}}(\theta_1, \varphi_1)\;
{_{-s_2}\tensor{Y}{^\star_{\ell,m}}}(\theta_2, \varphi_2)
=\;
{_{s_1}\tensor{Y}{_{\ell,s_2}}}(\beta, \alpha) \eem^{-\ii s_1 \gamma}\,,
\ee 
where the angles $(\alpha, \beta,\gamma)$ are defined through the implicit relation
\be
R_{\mathrm{E}}(\alpha, \beta, \gamma)=
R_{\mathrm{E}}(\varphi_1, \theta_1, 0)^{-1} 
R_{\mathrm{E}}(\varphi_2, \theta_2, 0)\,. 
\ee 
Here $R_\mathrm{E}(\al,\beta,\ga)$ is the rotation matrix with the Euler angles $\al$, $\beta$ and $\ga$. More precisely
\be
R_{\mathrm{E}}(\alpha, \beta, \gamma) =
\begin{pmatrix}
\cos \alpha \cos \beta \cos \gamma - \sin \alpha \sin \gamma &\ & - \cos \gamma  \sin \alpha -\cos \alpha  \cos \beta \sin \gamma  &\ & \cos \alpha \sin \beta  \\ 
 \cos \beta \cos \gamma \sin \alpha + \cos \alpha \sin \gamma&\ & \cos \alpha \cos \gamma - \cos \beta \sin \alpha \sin \gamma   &\ & \sin \alpha \sin \beta \\ 
- \cos \gamma \sin \beta  &\ & \sin \beta \sin \gamma  &\ & \cos \beta  
\end{pmatrix}\,.
\ee 
Explicit expressions of the Spin Weighted Spherical Harmonics for $s=0,1,2$ and $\ell\leq2$ are given in Tables \ref{tab:l1} and \ref{tab:l2}. Note that the remaining cases can be deduced from the conjugation relation given by Eq.~\eqref{eq:ylmconj}. 

We also introduce the auxiliary polynomials $\tilde{P}_{\ell}(\mu)$ and $\tilde{Q}_{\ell}(\mu)$ which will be useful later. For $\mu=\cos\theta$ they are defined as
\bea
  \frac{2\ell+1}{16\pi} \tilde{Q}_{\ell}(\mu) &\equiv&     - \sqrt{\frac{2\ell+1}{4\pi}}\; _{+2}\tensor{Y}{_{\ell,+2}}(\theta,\pi/2) ~=~
        - \sqrt{\frac{2\ell+1}{4\pi}}\;
        {_{-2}\tensor{Y}{_{\ell,-2}}(\theta,\pi/2)}  \,, \\
  \frac{2\ell+1}{16\pi} \tilde{P}_{\ell}(\mu)   &\equiv&    - \sqrt{\frac{2\ell+1}{4\pi}}\; _{+2}\tensor{Y}{_{\ell,-2}}(\theta,\pi/2) ~=~
        - \sqrt{\frac{2\ell+1}{4\pi}}\;
        _{+2}\tensor{Y}{_{\ell,-2}}(\theta, \pi/2)  \,. 
\eea
 From the orthonormality condition Eq.~\eqref{eq:yortho}, it is easy to see that
\be \label{eq:intPQ}
\int_{-1}^{+1}\; \tilde{P}_{\ell_1}(\mu)
\tilde{P}_{\ell_2}(\mu)\, \mathrm{d}\mu =
\int_{-1}^{+1}\; \tilde{Q}_{\ell_1}(\mu)
\tilde{Q}_{\ell_2}(\mu)\, \mathrm{d}\mu =
\frac{32}{2\ell_1 +1} \delta_{\ell_1, \ell_2}\,.
\ee 
The explicit expressions for these polynomials for $\ell=2,\dots,5$ are given in table \ref{tab:PQ}

\begin{center}
\begin{table}
\centering
\begin{tabular}{c|cc}
\hline  &&\\
$\ell$ & $\tilde{P}_\ell(\mu)$ & $\tilde{Q}_\ell(\mu)$ \\ &&\\ \hline \hline && \\ 
$2$ & $(\mu+1)^2$ & $(\mu-1)^2$ \\ && \\
 $3$  & $(\mu+1)^2(3\mu-2)$ & $(\mu-1)^2(3\mu+2)$ \\&& \\
$4$ & $(\mu+1)^2 (7\mu^2 -7\mu+1)$ & $(\mu-1)^2 (7\mu^2 +7\mu+1)$ \\ &&\\
$5$ & $(\mu+1)^2 (15\mu^3-18\mu^2+3\mu+1)$~~ &  ~~ $(\mu-1)^2 (15\mu^3+18\mu^2+3\mu-1)$\\ &&\\
\hline
\end{tabular}
\caption{The polynomials $\tilde{P}_\ell(\mu)$ and $\tilde{Q}_\ell(\mu)$. \label{tab:PQ}} \vspace{0.2cm}
\end{table}
\end{center}

\begin{center}
\begin{table}

\centering
\begin{tabular}{c|cc} 
\hline  &&\\
$m$    & $\tensor{Y}{_{1,m}}(\theta, \varphi)$ &  $_1\tensor{Y}{_{1,m}}(\theta, \varphi)$ \\ && \\ \hline \hline && \\
$-1$   & $\frac 12 \sqrt{\frac{3}{2\pi}} \eem^{-\ii \varphi} \sin \theta$ & $- \frac 14 \sqrt{\frac{3}{\pi}} \eem^{-\ii \varphi}  (1+\cos \theta)$   \\ && \\
$0$  & $\frac 12 \sqrt{\frac{3}{\pi}} \cos \theta$ & $\frac 12 \sqrt{\frac{3}{2\pi}} \sin \theta $ \\ &&  \\
$1$    & $- \frac 12 \sqrt{\frac{3}{2\pi}}\eem^{\ii \varphi} \sin \theta $ & $\frac 14 \sqrt{\frac{3}{\pi}}  \eem^{\ii \varphi}(-1+ \cos \theta )$  \\ && \\
\hline
\end{tabular}
\caption{Spherical Harmonics of Spin Weight $s=0,1$ and $\ell=1$ \label{tab:l1}}
\end{table}
\end{center}

\begin{center}
\begin{table}
\centering
\begin{tabular}{ c|ccc}
\hline  &&&\\
$m$    & $\tensor{Y}{_{2,m}}(\theta, \varphi)$ &  $_1\tensor{Y}{_{2,m}}(\theta, \varphi)$ &  $_2\tensor{Y}{_{2,m}}(\theta, \varphi)$ \\ &&& \\ \hline \hline  &&&\\
$-2$   & $\frac 14 \sqrt{\frac{15}{2\pi}} \eem^{-2\ii \varphi} \sin^2 \theta$ & $-\frac 14 \sqrt{\frac{5}{\pi}} \eem^{-2\ii \varphi} (1+\cos \theta)\sin \theta$ & $\frac{1}{8} \sqrt{\frac{5}{\pi}} \eem^{-2\ii \varphi} (1+\cos \theta)^2$\\ &&&\\
$-1$   & $\frac 12 \sqrt{\frac{15}{2\pi}}\eem^{-\ii \varphi} \sin \theta \cos \theta $ & $- \frac 14 \sqrt{\frac{5}{\pi}} \eem^{-\ii \varphi} (2 \cos^2 \theta + \cos \theta -1)$ &
$- \frac 14 \sqrt{\frac{5}{\pi}} \eem^{-\ii \varphi} \sin \theta   (1+\cos \theta)$  \\&&& \\
$0$    & $\frac 18 \sqrt{\frac{5}{\pi}} (1+3 \cos(2\theta))$ & $\frac 12 \sqrt{\frac{15}{2\pi}} \sin \theta \cos \theta$ & $\frac 14 \sqrt{\frac{15}{2\pi}} \sin^2 \theta $ \\ &&&\\
$1$    & $- \frac 12 \sqrt{\frac{15}{2\pi}}\eem^{\ii \varphi}  \sin \theta \cos \theta$ & $\frac 14 \sqrt{\frac{5}{\pi}}  \eem^{\ii \varphi}(2 \cos^2 \theta - \cos \theta -1)$ & $\frac 14 \sqrt{\frac{5}{\pi}} \eem^{\ii \varphi}\sin \theta (-1+\cos(\theta))$ \\&&& \\
$2$    & $\frac 14 \sqrt{\frac{15}{2\pi}} \eem^{2\ii \varphi} \sin^2 \theta$ & $\frac 24 \sqrt{\frac{5}{\pi}}\eem^{2\ii \varphi}  \sin \theta (-1+\cos \theta)$ & $\frac 18 \sqrt{\frac{5}{\pi}} \eem^{2\ii \varphi}(1-\cos \theta)^2$
\\ &&& \\ \hline
\end{tabular}
\caption{Spherical Harmonics of Spin Weight $s=0,1,2$ and $\ell=2$ \label{tab:l2}}
\end{table}
\end{center}

\subsection{Expression of the shear}
In this Appendix, we present useful relations involving spin Spherical Harmonics. More details can be found in~\cite{Durrer:2020fza, Seibert:2018}. This second reference is a very useful PhD thesis covering  the topic in depth. The interested reader is referred to it for further details. 

Let $(\bee_1, \bee_2)$ be an orthonormal basis on the sphere associated with the usual spherical coordinates $(\theta, \varphi)$. We define the $(+,-)$ basis
\be
\boldsymbol{e}_{\pm} = \frac{1}{\sqrt{2}} \left( \boldsymbol{e}_1 \mp \ii \boldsymbol{e}_2 \right)\,.
\ee
The spin raising and lowering operators are simply related to the covariant derivatives in directions $\boldsymbol{e}_{\pm}$,
\be
\boldsymbol{\nabla}_{\boldsymbol{e}_-} = - \frac{1}{\sqrt 2} \sd\,, \quad
\boldsymbol{\nabla}_{\boldsymbol{e}_+} = - \frac{1}{\sqrt 2} \sds\,.
\ee
With these identities, the relevant operators to compute the shear from the lensing potential are
\be 
\boldsymbol{\nabla}_1^2  - \boldsymbol{\nabla}_2^2 =
\frac 12 (\sd^2 + {\sds}^2)\,,
\ee 
and
\be 
\boldsymbol \nabla_1 \boldsymbol \nabla_2 =
- \frac{\ii}{4} (\sd^2 - {\sds}^2)\,,
\ee 
where it is assumed $\sd \sds = \sds \sd$, as in this context it acts on the scalar lensing potential $\phi$.   The definition of the shear in the $(\bee_1, \bee_2)$ basis is
\begin{align}
    \gamma_1 &= - \frac 12  (\boldsymbol{\nabla}_1^2  - \boldsymbol{\nabla}_2^2) \phi\,, \\
    \gamma_2 &= - \boldsymbol \nabla_1 \boldsymbol \nabla_2  \phi\,,
\end{align}
where $\phi$ is the lensing potential. This shows that the shear is a spin $2$ object. Using $\gamma^{\pm} = \gamma_1 \pm \ii \gamma_2$ and the relations given above, the shear in the $(+,-)$ basis is given by the slashed derivatives of the lensing potential as
\begin{align}
    \gamma^+ &= - \frac 12 {\sd}^2 \phi\,, \\
    \gamma^- &= -\frac 12 {\sds}^2 \phi\,.
\end{align}
Hence $\ga^+$ has helicity $+2$ while $\ga^-$ has helicity $-2$.
Using the standard decomposition for the lensing potential
\be
\phi(\boldsymbol n,z) = \sum_{\ell,m} \phi_{\ell,m}(z) Y_{\ell,m}(\boldsymbol n)\,,
\ee
and the squared raising/lowering operators given in Eq~\eqref{eq:ytwo}, one obtains the decomposition of the shear in the $(+,-)$ basis as
\begin{align} 
\label{eq:gammap}
\gamma^+(\boldsymbol n,z) &= -\frac 12 \sum_{\ell=2,m} \phi_{\ell,m}(z) \rll  \; _{+2}{Y}{_{\ell,m}}(\boldsymbol n)\,, \\
\label{eq:gammam}
\gamma^-(\boldsymbol n,z) &= -\frac 12 \sum_{\ell=2,m} \phi_{\ell,m}(z) \rll\;  _{-2}{Y}{_{\ell,m}}(\boldsymbol n)\,.
\end{align}
The complex numbers $\phi_{\ell,m}(z)$ are random variables whose expectation values define the angular power spectrum of the lensing potential,
\be
\langle \phi_{\ell_1, m_1}(z_1) \phi^\star_{\ell_2,m_2}(z_2) \rangle
= C_{\ell}(z_1,z_2) \delta_{\ell_1, \ell_2} \delta_{m_1,m_2}\,.
\ee 
As the lensing potential $\phi(\boldsymbol n,z)$ is real. They satisfy 
\be \label{eq:realphi}
\phi_{\ell,m}^\star = (-1)^m \phi_{\ell,-m}\,.
\ee

\subsection{Correlation functions on the equator}
\label{sec:corequa}
In this Section, we compute the correlation functions of the shear in the $(+,-)$ and $(\bee_1, \bee_2)$ basis. Using the decomposition given by Eq.~\eqref{eq:gammap} and Eq.~\eqref{eq:gammam} we find for the correlation function of $\ga^+$ and $\ga^{-}$
\begin{align}
    \langle  \gamma^+(\boldsymbol n_1, z_1) \gamma^-(\boldsymbol n_2, z_2)\rangle  &=  
    \frac 14 \sum \langle \phi_{\ell_1, m_1}(z_1) \phi_{\ell_2, m_2}(z_2) \rangle  \nu_{\ell_1} \nu_{\ell_2}\;  _{+2}{Y}{_{\ell_1,m_1}}(\boldsymbol n_1)\;  _{-2}{Y}{_{\ell_2,m_2}}(\boldsymbol n_2) \\
    &= \frac 14 \sum C_{\ell}(z_1,z_2) \rll^2 (-1)^{m}\; _{+2}{Y}{_{\ell,m}}(\boldsymbol n_1) \;
    _{-2}{Y}{_{\ell,-m}}(\boldsymbol n_2) \\
    &= \frac 14  \sum C_{\ell}(z_1,z_2) \rll^2  \;
    _{+2}{Y}{_{\ell,m}}(\boldsymbol n_1) \;
   _{+2}{Y}{^\star_{\ell,m}}(\boldsymbol n_2)\,,
\end{align}
where we used the conjugation properties Eq.~\eqref{eq:realphi} and  Eq.~\eqref{eq:ylmconj}. Using the addition theorem Eq.~\eqref{eq:summation} (with $s_1=+2$ and $s_2=-2$), the sum over $m$ reads
\begin{align}
    \sum_m   {_{+2}{Y}{_{\ell,m}}(\boldsymbol n_1)} \;
    _{+2}Y^\star_{\ell,m}(\boldsymbol n_2)
    &= - \sqrt{\frac{2\ell+1}{4\pi}}\; _{+2}\tensor{Y}{_{\ell,-2}}(\varphi, \pi/2) \\
    &\equiv \frac{2\ell+1}{16\pi} \tilde{P}_{\ell}(\mu)\,,
\end{align}
Finally, the correlation function is given by
\be \label{eq:corrpm}
\langle  \gamma^+(\boldsymbol n_1, z_1) \gamma^-(\boldsymbol n_2, z_2)\rangle 
=
   \sum_{\ell} \frac{2\ell+1}{64\pi} \rll^2 C_{\ell}(z_1,z_2) \tilde{P}_\ell(\mu)\,.
\ee
The two other correlations can be obtained following exactly the same steps (the values of the Euler angles are the same), yielding
\begin{align} \label{eq:corrpp}
    \langle  \gamma^+(\boldsymbol n_1, z_1) \gamma^+(\boldsymbol n_2, z_2)\rangle =
    \langle  \gamma^-(\boldsymbol n_1, z_1) \gamma^-(\boldsymbol n_2, z_2)\rangle  =
    \sum_{\ell} \frac{2\ell+1}{64\pi} \rll^2 C_{\ell}(z_1,z_2) \tilde{Q}_\ell(\mu)\,.
\end{align}
Inverting the relations $\gamma^{\pm} = \gamma_1 \pm \ii \gamma_2$ yields
\begin{align}
    \gamma_1 &= \frac 12 (\gamma^+ + \gamma^-) \,, \\
    \gamma_2 &= \frac{\ii}{2}(\gamma^- - \gamma^+)\,.
\end{align}
Using the correlations given above by Eq.~\eqref{eq:corrpm} and Eq.~\eqref{eq:corrpp} yield
\begin{align}
\langle  \gamma_1(\boldsymbol n_1, z_1) \gamma_1(\boldsymbol n_2, z_2)\rangle &  =
\sum_{\ell} \frac{2\ell+1}{128\pi} \rll^2 C_{\ell}(z_1,z_2) (\tilde{P}_\ell(\mu)+ \tilde{Q}_\ell(\mu))\,, \\
\langle  \gamma_2(\boldsymbol n_1, z_1) \gamma_2(\boldsymbol n_2, z_2)\rangle &  =
\sum_{\ell} \frac{2\ell+1}{128\pi} \rll^2 C_{\ell}(z_1,z_2) (\tilde{P}_\ell(\mu)- \tilde{Q}_\ell(\mu))\,, \\
\langle  \gamma_1(\boldsymbol n_1, z_1) \gamma_2(\boldsymbol n_2, z_2)\rangle &  =0\,,
\end{align}
where the points $\boldsymbol n_1$ and  $\boldsymbol n_1$ lie on the equator and subtend and angle $\varphi$ with $\mu=\cos\varphi$.

\subsection{Invariant correlation functions}
\label{sec:invcorfun}
\begin{figure}[ht!]
\begin{center}
\includegraphics[width=0.47\textwidth]{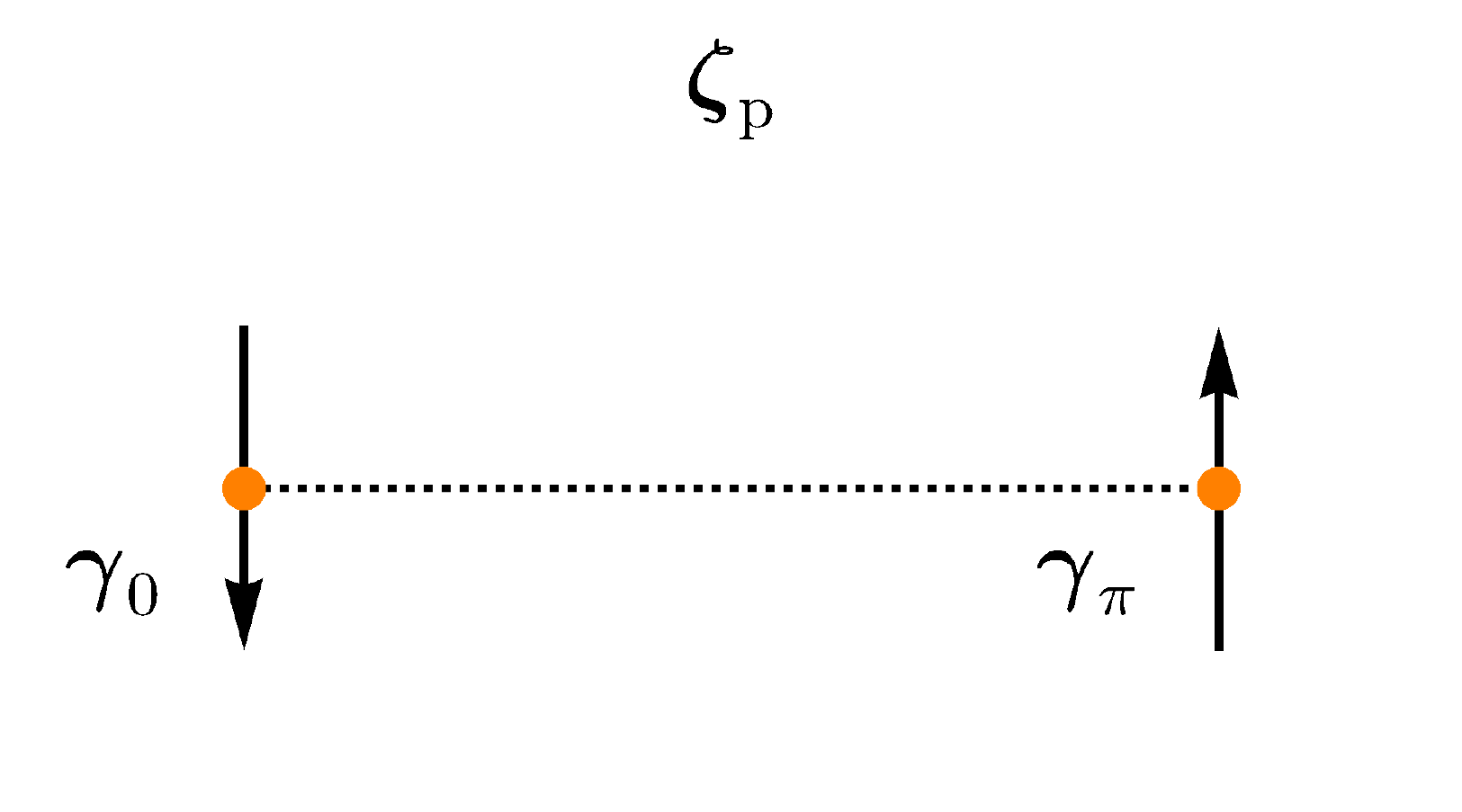}  \includegraphics[width=0.47\textwidth]{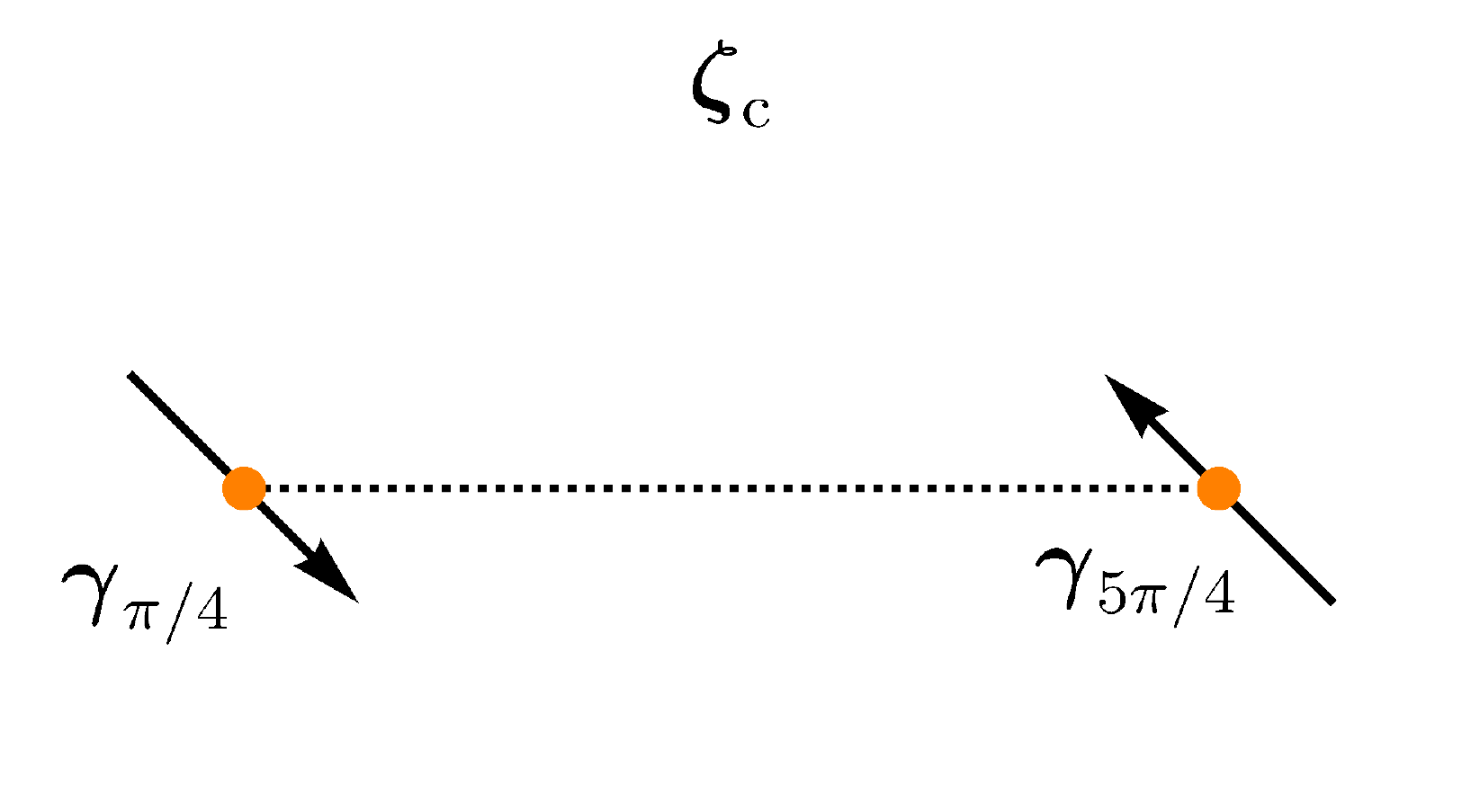}
\end{center}
\caption{\label{fig:zetacp} The correlation functions $\zp$ and $\zc$. Both galaxies are located on the 'Equator' - represented by the dotted lines - which defines a preferred system of coordinates. The angles between the direction of the shear and the connecting line are indicated. For a fixed separation angle, these correlations are intrinsically given and do not depend on the coordinate system.}
\end{figure}

Here, we  compute the shear and its correlation functions in a coordinate invariant way, see for example \cite{Ghosh:2018}. Let $(\theta, \varphi)$ be the spherical coordinates and $(\bee_1, \bee_2)$ the associated orthonormal frame. With such a basis, the shear is a 2-tensor of the form 
\be
\boldsymbol{\Gamma}=
\begin{pmatrix}
-\gamma_1 & -\gamma_2 \\ 
-\gamma_2 & \gamma_1 
\end{pmatrix}\,.
\ee 
For a generic tangent vector $\boldsymbol{e}=(\cos \alpha, \sin \alpha)$ in the $(\bee_1, \bee_2)$, the shear in direction $\boldsymbol{e}$ is defined as
\be
\gamma_{\alpha} \equiv \gamma_{\boldsymbol{e}} \equiv e^a e^b \Gamma_{ab} = - \gamma_1 \cos(2\alpha) - \gamma_2 \sin(2 \alpha)\,.
\ee 
It is clear from the definition that $\gamma_{1,2}$ and the angle $\alpha$ do depend on the coordinate system. However, for a fixed (physically defined) vector $\boldsymbol{e}$, the shear in  direction $\boldsymbol{e}$,  $\gamma_{\boldsymbol{e}}$ does not depend on the coordinates, which makes this quantity a good candidate to study correlation functions. For two galaxies located at $(\boldsymbol n_1, z_1)$ and $(\boldsymbol n_2, z_2)$, we can define the geodesic joining them to be the equator of our system of coordinates. As this process does not depend on the  coordinates and is well-defined for every pair of galaxies, the result that follows is also coordinate independent. From this construction, we define the two invariant correlation functions
\begin{align}
    \zp(\mu, z_1,z_2) &=  \langle \gamma_{0}(\boldsymbol n_1, z_1) \gamma_{\pi}(\boldsymbol n_2, z_2) \rangle  = \langle \gamma_{1}(\boldsymbol n_1, z_1) \gamma_{1}(\boldsymbol n_2,z_2) \rangle \,,  \\
   \zc(\mu,z_1,z_2) &=  \langle \gamma_{-\pi/4}(\boldsymbol n_1, z_1) \gamma_{3\pi/4}(\boldsymbol n_2, z_2)\rangle  = \langle \gamma_{2}(\boldsymbol n_1, z_1)\gamma_{2}(\boldsymbol n_1, z_1)\rangle \,,
\end{align}
with $\mu = \boldsymbol n_1 \cdot \boldsymbol n_2 = \cos \varphi$. The last equality is valid in the preferred system of coordinates, where both galaxies lie on the equator. An illustration of this definition is shown in Fig.~\ref{fig:zetacp}. Using the results of Sec.~\ref{sec:corequa} yields
\begin{align}
\label{eq:zpexpr}
\zp(\mu,z,z^\prime) &= \sum_\ell \frac{2\ell+1}{128\pi} C_\ell(z,z^\prime) \rll^2 ( \tilde{P}_{\ell}(\mu) + \tilde{Q}_\ell(\mu)) \,, \\
\label{eq:zcexpr}
\zc(\mu,z,z^\prime) &= \sum_\ell \frac{2\ell+1}{128\pi} C_\ell(z,z^\prime) \rll^2 ( \tilde{P}_{\ell}(\mu) - \tilde{Q}_\ell(\mu)) \,.
\end{align}
Note that the sums start at $\ell=2$. 
Defining,
\begin{align}
        \label{eq:zetap}
        \zeta_{\pm} &= \frac{1}{2} (\zp \pm \zc)\,, 
\end{align}
and using the orthogonality properties of the polynomials $\tilde{P}_\ell$ and $\tilde{Q}_\ell$ given in Eq.\eqref{eq:intPQ}, we have
\begin{align}
    \label{eq:zetapexpr}
    \int_{-1}^{+1} \zeta_+(\mu,z_1,z_2) \tilde{P}_\ell(\mu)\, \mathrm{d}\mu &= \frac{1}{4\pi} C_\ell(z_1,z_2) \rll^2 \,, \\
    \label{eq:zetamexpr}
    \int_{-1}^{+1} \zeta_-(\mu,z_1,z_2) \tilde{Q}_\ell(\mu)\, \mathrm{d}\mu &= \frac{1}{4\pi} C_\ell(z_1,z_2) \rll^2 \,.
\end{align}
Note that the relations Eq.~\eqref{eq:zetapexpr} and Eq.~\eqref{eq:zetamexpr} relate only coordinate independent observables. The correlation functions $\zeta_{\pm}$ can be estimated by observations as explained in the main text. Via \eqref{eq:zetapexpr} we can then use them to estimate the lensing power spectrum $C_{\ell}$. 

\section{Error estimation \label{app:error}}
\subsection{Best estimator}
Given $X_j$ measurements of an observable $X$, each of them with error $\delta X_j = \tau_j X_j$ ($\tau$ is the relative error). We want to construct an estimator for $X$. We define 
\be 
\hat{X} = \sum w_j X_j\,, \qquad \sum w_j=1n\,.
\ee 
In order to obtain the best possible estimator for $X$
We want to choose the weights $w_j $ which yield the highest signal-to-noise ratio (SNR). We claim
\be  \label{eq:ansatzalpha}
w_j = \frac{1}{Z} \frac{1}{X_j \tau_j^2}\qquad \mbox{where} \quad Z = \sum \frac{1}{X_j \tau_j^2}\,.
\ee 
To see that this is the best choice, we note that
the error on the estimator is given by
\be
N^2 = \sum w_j^2 \delta X_j^2 = \sum w_j^2 \tau_j^2 X_j^2\,.
\ee 
The square of the SNR which we want to maximise is the quantity
\be
A = \frac{\hat{X}^2}{N^2}.
\ee 
Using the Ansatz \eqref{eq:ansatzalpha}, one can verify directly that this is choice of the weights gives
\be 
\frac{\partial A}{\partial w_i} =0\,,
\ee 
and it is the only zero of the gradient of $A$ (with positive weights which sum up to 1) and it is a minimum. 
Hence the $w_i$ given above are the best choice if one wants to maximize the SNR of an observable. The constant $Z$ is determined by the requirement that
$$
\sum w_j =1 \,.
$$
Computing $A$ explicitly one finds
the well known result
\be \label{eq:SNR}
{\rm SNR} = \sqrt{A} = \sqrt{\sum \frac{1}{\tau_j^2}}\,.
\ee 

\subsection{Specific example}
If we consider our estimator of the correlation function, $\zeta_\pm(\mu,z_1,z_2)$ and denote the value obtained from two pairs of galaxies by $\hat{\zeta}_j$ and the error by $\delta \hat{\zeta}_j$, then we find
\be 
A = \sum \left( \frac{\hat{\zeta}_j}{\delta \hat{\zeta}_j} \right)^2\,,
\ee 
and the optimal estimator for $\zeta_\pm(\mu,z_1,z_2)$ is
\be 
\hat{\zeta}(\mu,z_1,z_2) = \sum \frac{1}{Z} \left( \frac{\hat{\zeta}_j}{\delta \hat{\zeta}_j}\right)^2\,,
\ee
with 
\be 
Z = \sum \frac{\hat{\zeta}_j}{(\delta \hat{\zeta}_j)^2}\,.
\ee

\subsection{Counting the pairs of galaxies}
Here we want to count the number of pairs of galaxies which can be used to estimate $\xi_{\pm}(\mu,z_1,z_2)$. For this we need to estimate  the number of galaxies  with fixed opening angle $\varphi$, $\mu=\cos\varphi$. We suppose that we have pixels of angular aperture $\delta \theta$. The solid angle of a cone with this opening angle is at lowest order
\be
\delta \Omega = \delta \theta^2 \pi\,.
\ee 
Let us set the first pixel at the North Pole. We want to count the number of pixels whose center is at an angle $\varphi\pm \delta \theta/2$ from this first pixel. The solid volume of these pixels is 
\be 
\delta \Omega_{\varphi} = 2\pi \int_{\varphi-\delta \theta/2}^{\varphi + \delta \theta/2}\; \sin \theta \, \mathrm{d} \theta =
2 \pi \delta \theta \sin \varphi\,.
\ee 
Here we assume that the full ring with angle $\varphi$ around the first pixel is observed. For incomplete sky coverage this is not true for all values of $\varphi$, but we neglect this in our treatment and take the sky coverage into account as an overall factor $\fsky$ which denotes the fraction of the sky covered by the survey.
Hence, the number of pixels forming such an angle with the original pixel is given by
\be
N(\varphi) =  \frac{\delta \Omega_{\varphi}}{\delta \Omega} 
= \frac{2 \sin(\varphi)}{\delta \theta}\,.
\ee
We also need the total number of pixels which we can choose as our first pixel, given by 
\be 
N_{\rm tot} = \fsky \frac{4\pi}{\delta \Omega} = \frac{4\fsky}{\delta \theta^2}\,.
\ee 
Here we have introduced $ f_{\rm sky}$, the observed sky fraction.
The total number of couples separated by an angle $\varphi$ is
\be
\Nc(\varphi) = N_{\rm tot} \times N(\varphi) = \frac{8 \fsky \sin \varphi}{\delta \theta^3}\,.
\ee
If we consider auto-correlations, $z_1=z_2$, this number has to be be divided by $2$ due to symmetry. Let us now denote the number of galaxies in a pixel at redshift $z$ by $\Ng(z)$. For a given pair of pixels at $z_1$ and $z_2$, one can choose $\Ng(z_1) \Ng(z_2)$ pairs of galaxies. Hence, the total number of pairs of galaxies which we can consider for the estimator  $\hat\zeta_\pm(\varphi,z_1,z_2)$ is
\be
\Np(\varphi,z_1,z_2) =  \Ng(z_1) \Ng(z_2) \Nc(\varphi) \,,
\ee 
To compute an estimator $\hat{\zeta}_\pm$, we need $4$ galaxies, or $2$ different pairs. The number of estimators we can form is therefore
\be \label{eq:nefinal}
\Ne(\varphi,z_1,z_2) = 
\frac{\Np(\varphi,z_1,z_2) (\Np(\varphi,z_1,z_2)-1)}{2} 
\simeq \frac{1}{2}\left(\Ng(z_1) \Ng(z_2) \frac{8 \fsky \sin \varphi}{\delta \theta^3}\right)^2\,.
\ee 
The division by $2$ of $\Nc$ becomes a division by $4$ of $\Ne$ if we consider auto-correlations, $z_1=z_2$.

\bibliographystyle{JHEP}
\bibliography{refpap}


\end{document}